\documentclass[aps, prd, reprint,showpacs,nofootinbib]{revtex4-1}

\usepackage{amssymb, amsmath, braket, nicefrac, graphicx} 
\usepackage[tight]{subfigure}
\usepackage{relsize}

\newcommand{\numu}{\ensuremath{\nu_\mu } }                   
\newcommand{\nue}{\ensuremath{\nu_e } }                      

\begin{document}

\title{\bf Transition probability of perturbative form for 
       $\nu_{\mu} \rightarrow \nu_{e}$ oscillations in matter of constant density}

\author{W. Anthony \surname{Mann}}
\author{Tomas   \surname{Kafka}}
\author{Jacob   \surname{Schneps}}
\author{Ozgur  \surname{Altinok}}

\affiliation{Tufts University, Medford, MA 02155}


\begin{abstract}
We give a convenient expression for the appearance probability $\mathcal{P}(\nu_{\mu} \rightarrow \nu_{e})$ describing neutrino
oscillations in matter of constant density,  derived using textbook quantum mechanics stratagems.   Our formulation retains the clarity of an 
expansion in $\alpha =  {\Delta m_{21}^2}/{\Delta m_{31}^2}$ exhibited by the popular Cervera {\it et al.}  formula  [Nucl. Phys. B {\bf 579}, 17 (2000)]
while enabling more accurate evaluation of oscillations over terrestrial baselines.
\end{abstract}

\maketitle

\section{Introduction} \vspace{-9pt}

Analytic forms for flavor transition probabilities of neutrino oscillations in matter can facilitate studies of measurement sensitivity afforded by
proposed new experimental facilities \cite{ref:Barger-2002, ref:Huber-Winter-2003, ref:DGR-PRL, ref:Ballett-Pascoli, ref:Bishai}.
Of particular current interest is the transition probability for subdominant $\nue$ appearance, 
$\numu \rightarrow \nue$, using $\numu$ beams 
propagating over long baselines through terrestrial matter.   Indeed, the size of the neutrino mixing angle $\theta_{13}$ has been
dramatically clarified during the past year as the result of a number of recent experimental measurements
\cite{ref:T2K-big-theta13, ref:MINOS-improved-nue, ref:Double-Chooz, ref:Daya-Bay, ref:RENO}.

A determination which is representative of the new level of precision is reported by the Daya Bay reactor experiment:
$\sin^2 2\theta_{13} = 0.092 \pm 0.016$ (stat) $\pm 0.005$ (sys).    With $\theta_{13} \sim 9^o$, the study of CP violation and mass 
hierarchy in the neutrino sector can proceed in earnest.  
Analytic investigations in this new era can benefit from the availability of convenient analytic forms for
$\mathcal{P}(\numu \rightarrow \nue)$ which are accurate to within a few percent for neutrino baselines 
through the Earth's mantle.    This work provides such a probability expression for $\numu \rightarrow \nue$
oscillations in a constant-density matter field.

A number of exact derivations for neutrino propagation 
with oscillations among three active flavors in constant-density matter 
have appeared in the literature over the past several decades \cite{ref:Zaglauer-Schwarzer,  ref:Ohlsson-JMP, ref:Ohlsson-PL, ref:Xing-PL, ref:Ohlsson-PS}.  
In general these formulations are complicated and do not readily yield insights.    A degree of clarity is achieved with the exact probability expressions
of Kimura, Takamura, and Yokomakura in which matter effects are disentangled from CP violation effects  \cite{ref:Kimura-PL, ref:Kimura-PRD}.   Their
formulation has been extended to include, for example, nonstandard interaction matter effects \cite{ref:Yasuda-KTY, ref:Ribeiro-KTY}.
Nevertheless, a desire for more transparent formulations has led to the development of
various approximation expansions;  a review with comparisons can be found in Ref. \cite{ref:Akhmedov}.
A treatment which incorporates the magnitude of $\theta_{13}$ as recently measured into a perturbative framework
is presented in~\cite{ref:Asano-Minakata}.

For the $\nue$ appearance probability $P(\numu \rightarrow \nue) $
the three-term formula of Cervera {\it et al.}~\cite{ref:Cervera} (see also \cite{ref:Freund}) 
is frequently utilized in analytic studies.   This formula has the form of a perturbative expansion in terms of the small
mass hierarchy ratio $\alpha \equiv \Delta m^2_{21}/\Delta m^2_{31} \simeq 1/32$.
For propagation through the Earth's crust such as occurs with the T2K (295 km), MINOS (735 km), and NO$\nu$A (810 km) baselines,
the formula of Ref.~\cite{ref:Cervera} is adequate for most purposes.     However for terrestrial baselines which exceed the proposed LBNE
baseline of 1300 km,  such as the ``bimagic" 2540 km baseline~\cite{ref:DGR-PRL} and the ``magic" 7500 km baseline~\cite{ref:Barger-2002, ref:Huber-Winter-2003}
which have significant pathlength through the Earth's mantle, more accurate formulations are desirable.
The Cervera {\it et al.} formula can be written as follows:
\smallskip
\begin{equation}  \label{eq:Cervera-three-terms}
\begin{split}
  \mathcal{P}&_{approx}(\nu_\mu \rightarrow \nu_e) \simeq  \\
& \sin^2 2 \theta_{13} ~s_{23}^2 \cdot \frac{  \sin^2((1-A)\Delta)}{(1-A)^2}  \\
 + &~\alpha \sin 2 \theta_{13} ~c_{13}\sin 2\theta_{12} \sin 2\theta_{23}  
       \cdot \frac{\sin(A \Delta)}{A} \cdot \\
&\frac{\sin((1-A)\Delta)}{(1-A)} \cdot  [ \cos \delta_{CP}   \cos \Delta -  \sin\delta_{CP}  \sin\Delta ]  \\
 + &~\alpha ^2  c_{23}^2  \sin^2 2 \theta_{12}  \cdot  \frac{\sin^2(A \Delta)}{A^2}.
\end{split}
\end{equation}
\noindent
In the above expression and throughout this paper, we use $\Delta \equiv   \Delta m_{31}^2 \ell / (4 E_\nu)$ for
the atmospheric oscillation phase at baseline $\ell$.    The symbol $A$ refers to the matter potential $A \equiv
\pm (2\sqrt{2} G_{F} n_{e} E_{\nu})/ \Delta m_{31}^2$ where $G_{F}$ is the Fermi coupling constant and $n_{e}$ is
the electron density in matter.   The sign of $A$ is determined by the sign of $ \Delta m_{31}^2$ and choice of
neutrino or antineutrino propagation.     As a matter of convention, the formalism of this work refers to neutrino
propagation (the + sign) and assumes the normal mass hierarchy for the neutrino mass eigenstates ($ \Delta m_{31}^2$ positive).   
We also use the compact notations $s_{ij} \equiv \sin\theta_{ij}$ and $c_{ij} \equiv \cos\theta_{ij}$ with $i,j$ = 1,2,3.

We present a formulation of the $\nue$ appearance probability
$\mathcal{P}(\nu_\mu \rightarrow \nu_e)$ which retains the convenient
perturbative form of Eq. \eqref{eq:Cervera-three-terms} in the leading 
three terms of a six-term expansion.  Our derivation uses conventional
non-relativistic quantum mechanics methods to construct an approximate
form for the time evolution operator for neutrino states in flavor basis.
Amplitudes of useful precision are thereby implied for all  
oscillation transitions accessible to neutrinos of three active 
flavors.   In this work however we focus upon $\numu \rightarrow \nue$
oscillations. Our approach is amenable to augmentations such as 
obtaining oscillation amplitudes with inclusion of selected
non-standard interaction (NSI) matter potentials, but such developments
are left for a future work.

\section{Outline} \vspace{-9pt}

The paper proceeds as follows:  We define some convenient notations and
proceed straightaway in Sec. \ref{sec:Exact} to report our result, stating the
three amplitudes (Sec. \ref{sec:Conv}) and the  consequent six-term oscillation
probability (Sec. \ref{sec:ThreeTerms}) which comprises our formulation
for $\mathcal{P}(\nu_\mu \rightarrow \nu_e)$ in a constant density medium.
We then show that our leading terms have resemblances
to the probability terms of Eq. \eqref{eq:Cervera-three-terms}.
The remaining three terms of our expression entail relatively small contributions to
$\mathcal{P}(\nu_\mu \rightarrow \nu_e)$.
The extent of variations is illustrated using plots of our six-term probability versus
Eq. \eqref{eq:Cervera-three-terms}
for $E_{\nu}$ between one and ten GeV
at the proposed LBNE, the bimagic, and the magic baselines (Sec. \ref{sec:SixTerms}).

The derivation of our 
$\mathcal{P}(\nu_\mu \rightarrow \nu_e)$ formula is given in Sec. \ref{sec:Derivation}.
In brief, the conventional three-flavor Hamiltonian in flavor basis
is transformed to propagation basis, whereupon it is re-phased and then separated
into an ``unperturbed" piece plus an interaction potential of relatively small coupling
strength (all elements proportional to $\sin 2\theta_{12} \cdot \alpha$).  
The latter is used to set up an
interaction picture and the time evolution operator for neutrino states is constructed in a 
heuristic way in that picture.  A key step is the exponentiation of the interaction potential into
 $\exp(-i \hat V_{I}(t))$; the latter can be expressed as a matrix identity
 which has similar form to the well-known identity for
 $exp(-i \hat J_{y} \theta)$ for $j$ = 1 angular momentum.   Our
 heuristic form of the evolution operator is then transformed back into 
propagation basis and finally into flavor basis, yielding our approximate solution for
 $\hat U^{(\alpha)}$.    In that basis, its matrix 
elements coincide with the various three-flavor neutrino oscillation
amplitudes.   The $\nue$ appearance amplitude is given by
$ \hat U^{(\alpha)}_{12}$.

\vspace{-9pt}
\section{Formulae for $\nu_{\mu} \rightarrow \nu_{e}$ oscillations} 
\label{sec:Exact}

\subsection{Conventions and notations} \vspace{-9pt}
\label{sec:Conv}

In expressions to follow we refer to the vacuum oscillation length
$\ell_v = {E_\nu} / {\Delta m_{31}^2}$, and we write
\begin{equation}\label{eq:Define_Delta}
    \Delta =  \Delta m_{31}^2 \ell / (4 E_\nu) =  \frac{\ell}{4\ell_v} ~~~ \text{and} ~~~V_e \equiv \frac{A}{2\ell_v} ~.
\end{equation}
Convenient quantities are the neutrino mass hierarchy ratio $\alpha \simeq 1/32$ and
the scaled forms for the hierarchy parameter $\alpha'$ and $\alpha''$  defined as
\begin{eqnarray}
\label{eq:alpha}
 \alpha' &\equiv& \sin 2\theta_{12} \cdot \alpha~, ~~\text{and} \\
 \alpha'' &\equiv& (1-3 c_{12}^2) \cdot \alpha~. \nonumber
\end{eqnarray}
In our expressions, the mixing strengths involving $\theta_{13}$ are often accompanied by the factor $(1 - s_{12}^2\alpha)$, hence we define
\begin{eqnarray}
 \sin 2 \tilde\theta_{13} &=&(1 - s_{12}^2\alpha) \cdot  \sin 2 \theta_{13} , ~~\text{and} \nonumber \\
 \cos 2 \tilde\theta_{13} &=& (1 - s_{12}^2\alpha) \cdot \cos 2 \theta_{13} .
\end{eqnarray}

\subsection{Oscillation amplitude of three terms}
\label{sec:ThreeTerms}

The amplitude for $\nue$ appearance at baseline $\ell$ from an initial $\numu$ beam which propagates through matter 
of constant density, can be expressed as a sum of three terms:
\begin{equation}\label{eq:define_amplitude}
  \mathcal{A}(\nu_\mu \rightarrow \nu_e) = T_1 + T_2 + T_3 ~~.
\end{equation}
The individual $T_{i}$ terms are the following:
\begin{equation}\label{eq:T_1_final}
  T_1 = (-i)  \sin 2\tilde\theta_{13}  ~s_{23} \cdot \frac{ \sin (N\ell) }{4 \ell_v N} \cdot e^{-i\delta_{CP}},
\end{equation}
\begin{equation}\label{eq:T_2_final}
  T_2 = (-i)~ c_{13} c_{23} \cdot \sin(\eta \ell) \cdot e^{iG\ell},
\end{equation}
and
\begin{eqnarray}
\label{eq:T_3_final}
  T_3 = &\sin 2\theta_{13}~ s_{23}\cdot \sin^2 \left( \frac{\eta \ell}{2} \right) \cdot ~~~~~~~~~~\nonumber\\
        & \left[ \cos(N\ell) + i F_{A}  \frac{\sin(N\ell)}{4 \ell_v N} \right]  \cdot e^{-i\delta_{CP}}.
\end{eqnarray}
The factor $N$ which appears in oscillation phases 
in Eqs. (\ref{eq:T_1_final}) and (\ref{eq:T_3_final}) is defined by
\begin{equation}\label{eq:Define-N}
  N \equiv \frac{1}{4\ell_v} \left[ (\sin 2\tilde\theta_{13})^2 + (\cos 2\tilde\theta_{13} - A)^2 \right]^{\frac{1}{2}} ~.
\end{equation}
In Eqs.  (\ref{eq:T_2_final}) and (\ref{eq:T_3_final}) we use
\begin{equation}\label{eq:Define-eta_and_G}
\begin{split}
  \eta \equiv \frac{\alpha'}{4\ell_v}\;\; ,  ~~~~
  G \equiv \frac{1}{4\ell_v} \left[(1+A)+\alpha''\right] ~~.
\end{split}
\end{equation}
The quantity $F_{A}$ appearing in Eq.~(\ref{eq:T_3_final}) is
\begin{equation}\label{eq:Define-FA}
\begin{split}
  F_A \equiv \left[ c_{13}^2(1-s_{12}^2\alpha) - (\cos 2\tilde\theta_{13}-A) \right]   \simeq A .
\end{split}
\end{equation}

The variables designated in \eqref{eq:Define-N}, \eqref{eq:Define-eta_and_G}, and \eqref{eq:Define-FA} are ones
which arise naturally in the derivation of Sec.~\ref{sec:Derivation}.   To facilitate comparison with Eq. \eqref{eq:Cervera-three-terms} we
also define as convenient quantities $D$ and $\Delta'$:
\begin{equation}
\label{eq:Define-D}
D \equiv 4\ell_{v} N  \simeq |1-A|~, 
\end{equation}
\begin{equation} \label{eq:define_Delta-prime}
 \Delta' \equiv G \ell = \Delta \left[ (1+A) + \alpha'' \right] ~.
\end{equation}
Thus the oscillation phases $N\ell$ and $\eta \ell$ can be written as 
$(D/4\ell_{v})\ell = D \Delta$ and as $(\alpha'/4 \ell_{v}) \ell = \alpha' \Delta$ respectively.

\vspace{-9pt}
\subsection{Oscillation probability of six terms}
\label{sec:SixTerms}

The $\nu_\mu \rightarrow \nu_e$ oscillation probability is constructed from
 $|\mathcal{A}(\nu_\mu \rightarrow \nu_e)|^2$.
Referring to Eq. (\ref{eq:define_amplitude}), we express the result as a sum of six terms:
\begin{eqnarray} 
\label{eq:probability}
  \mathcal{P}(\nu_\mu \rightarrow \nu_e)
    = ~&|&T_1|^2 + (T_1 T_2^* + T_1^* T_2) + \nonumber \\
      &|&T_2|^2 + (T_1 T_3^* + T_1^* T_3) + \nonumber \\
      &(&T_2 T_3^* + T_2^* T_3) + |T_3|^2  ~.
\end{eqnarray}
We proceed to construct the individual probability terms which appear in Eq. (\ref{eq:probability}).   
Squaring the amplitude of Eq. \eqref{eq:T_1_final} we obtain
\begin{equation}  \label{eq:T_1_squared}
  |T_1|^2 = (\sin 2 \tilde\theta_{13})^2 ~s_{23}^2 \cdot \frac{  \sin^2(D \Delta)}{D^2} ~.
\end{equation}
Similarly, we obtain from Eq. \eqref{eq:T_2_final} 
\begin{equation}  \label{eq:T_2_squared}
  |T_2|^2 = c_{13}^2 c_{23}^2 \cdot \sin^2(\eta \ell)
          = c_{13}^2 c_{23}^2 \cdot \sin^2 (\alpha' \Delta).
\end{equation}

\smallskip
Considering the $T_1  T_2$ cross terms, we write
\begin{equation}
\begin{split}
  (T_1 T_2^* + &T_1^* T_2) = \\
    &\sin 2 \tilde\theta_{13}  ~c_{13} s_{23} c_{23} 
       \cdot \sin(\eta\ell) \cdot  \frac{\sin (N\ell)}{4 \ell_v N} \cdot \\
  &  \left[ e^{-i(G\ell+\delta_{CP})} + e^{i(G\ell+\delta_{CP})} \right].
\end{split}
\end{equation}
Upon introducing the notations of \eqref{eq:Define-D} and \eqref{eq:define_Delta-prime} 
and reducing the bracket expression, we obtain
\begin{equation} \label{eq:T1 -T 2-cross-final}
\begin{split}
  (T_1 T_2^*& + T_1^* T_2) = \\
    &\sin 2 \tilde\theta_{13} ~c_{13}\sin 2\theta_{23}   
       \cdot \sin(\alpha' \Delta)  \cdot \frac{\sin(D \Delta)}{D} ~\cdot \\
    &[ \cos \Delta' \cdot \cos \delta_{CP} - \sin \Delta' \cdot \sin \delta_{CP} ].
\end{split}
\end{equation}

\smallskip
There are two more sets of cross terms to consider. We first determine $(T_1 T_3^* + T_1^* T_3)$:
\begin{equation} \label{eq:T_13_cross}
\begin{split}
  (T_1 T_3^* + T_1^* T_3)
    = &-2  \sin 2 \theta_{13}  \sin 2 \tilde \theta_{13}  s_{23}^2 ~ F_A \\
      &\cdot \sin^2 \left(\frac{\alpha' \Delta}{2} \right) \cdot \frac{ \sin^2(D \Delta)}{D^2}.
\end{split}
\end{equation}

The remaining cross term is $(T_2 T_3^* + T_2^* T_3)$.
Now $T_2 T_3^*$ is
\begin{equation} \label{eq:T_2_T_3s}
\begin{split}
  T_2 T_3^*
    &= (-i) ~ \sin 2 \theta_{13}~ c_{13} c_{23} s_{23} 
       \cdot \sin(\eta\ell) \cdot \sin^2 \left(\frac{\eta\ell}{2}\right) \\
    &\; \cdot \left[ \cos(N\ell) - i~F_{A} \cdot \frac{\sin(N\ell)}{D} \right] \cdot e^{+i(G\ell+\delta_{CP})}.
\end{split}
\end{equation}
Adding $T_2^* T_3$ to $T_2 T_3^*$ and extracting the common factors, we obtain
\begin{equation}\label{eq:T2T3cross} 
\begin{split}
 (&T_2 T_3^* + T_2^* T_3) = \\
   &\left[ \frac{1}{2} \sin 2\theta_{13} ~c_{13} \sin 2\theta_{23}   \cdot \sin(\eta\ell)   
             \cdot \sin^2 \left(\frac{\eta\ell}{2}\right) \right] \cdot \\
 &\left\{ (-i) e^{i(G\ell+\delta_{CP})} \left[ \cos(N\ell) - i F_A~\frac{\sin(N\ell)}{D} \right] + ~\text{c.c.} ~\right\}. 
\end{split}
\end{equation}
 Equation (\ref{eq:T2T3cross}) reduces to
\begin{equation}\label{eq:T_23_cross}
\begin{split}
  (&T_2 T_3^* + T_2^* T_3) = \\
 & \sin 2\theta_{13} ~c_{13} \sin 2 \theta_{23}  \cdot \sin(\alpha' \Delta) \cdot
       \sin^2 \left( \frac{\alpha' \Delta}{2} \right) \cdot \\
          &\left\{ \cos(D \Delta) \cdot \sin(\Delta' + \delta_{CP})
         - F_A\frac{ \sin(D \Delta)}{D} \cdot \cos(\Delta'+\delta_{CP}) \right\}  .
\end{split}
\end{equation}

The final term is $|T_3|^2$. Referring to Eq. (\ref{eq:T_3_final}) we obtain
\begin{equation} \label{eq:T_3_squared}
\begin{split}
  |T_3|^2
    =& ~ \sin^2 2\theta_{13} ~s_{23}^2\cdot \sin^4 \left(\frac{\alpha' \Delta}{2} \right) \cdot \\
     &  \left\{ \cos^2(D \Delta) + F_A^2  \cdot  \frac{\sin^2(D \Delta)}{D^2} \right\}. 
\end{split}
\end{equation}

The sum of the six probability terms 
of Eqs. \eqref{eq:T_1_squared}, \eqref{eq:T_2_squared},  \eqref{eq:T1 -T 2-cross-final}, 
\eqref{eq:T_13_cross}, \eqref{eq:T_23_cross}, and \eqref{eq:T_3_squared}  
comprise our rendering of the probability for $\nue$ appearance in an initial $\numu$ beam.
For convenience we state the total probability as a single expression in the conclusion (Sec.~\ref{sec:Conclusion}) of this work.

\subsection{Comparison to the Cervera {\it et al.} probability}
\label{sec:Comparison}

 In Eq. (\ref{eq:probability}) for the $\numu \rightarrow \nue$ oscillation probability, the three leading terms are
$ |T_1|^2, ( T_1 T_2^{*} + T_1^* T_2 )$, and $ |T_2|^2$, for which explicit expressions are given by
Eqs. \eqref{eq:T_1_squared},  \eqref{eq:T1 -T 2-cross-final}, and \eqref{eq:T_2_squared} respectively.    
These terms have a clear resemblance to the corresponding three terms 
of the formula given in Eq. \eqref{eq:Cervera-three-terms}, 
however our extended perturbative form gives rise to certain modifications.  In the two leading terms, 
our variables $\sin 2 \tilde \theta_{13}$ and $D$ replace 
$\sin 2 \theta_{13}$ and $(1-A)$ respectively.   For the second and third terms, 
the correspondence between our result versus the Cervera {\it et al.}  expression can be seen 
by invoking small angle approximations and by recognizing that $\Delta ' \simeq \Delta$ and
that $c^{2}_{13} \simeq 1.0$.

Equation (\ref{eq:probability}) contains three additional terms which arise from the presence 
of the very small $T_3$ amplitude.   These terms go beyond the level of accuracy intended 
with Eq.~\eqref{eq:Cervera-three-terms}.  
We find that these extra terms contribute amounts to the probability of less than 
one percent for terrestrial baselines accessible to accelerator-based oscillation experiments.  
Thus there is justification for neglecting these relatively complicated higher-order terms.   
The improved accuracy afforded by our formulation arises in the main 
with the refinements introduced into the three probability terms already present 
in the Cervera \textit{et al.} formula~\cite{ref:Cervera}, rather than in the extra terms.

\begin{figure}[!htb]
\includegraphics[width=8.3cm]{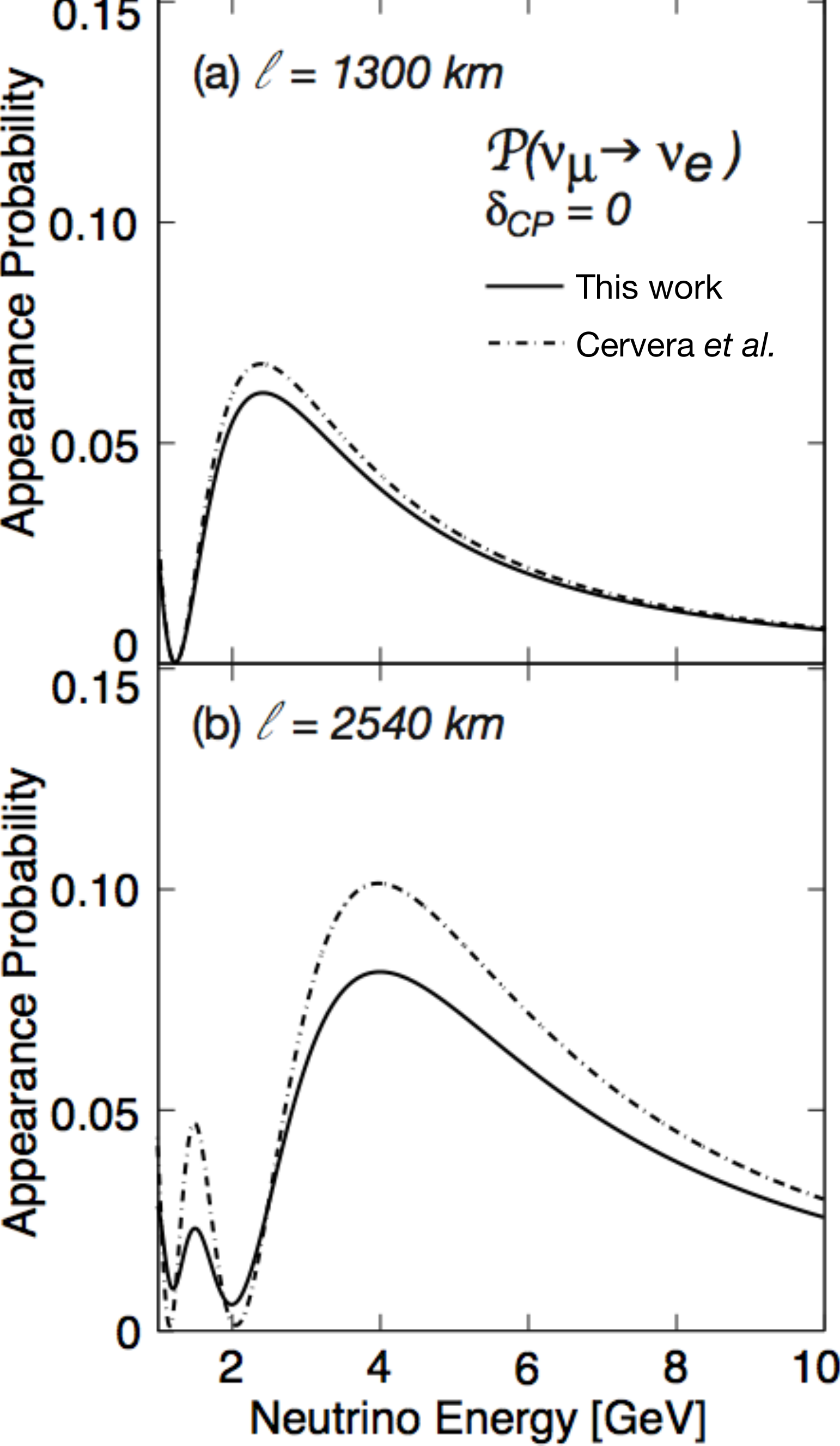}
\caption{ \small Probability for $\nue$ appearance from accelerator $\numu$ neutrino beams propagating in constant density matter, for the (a) LBNE and (b) bimagic baselines.   In each plot our six-term transition probability (solid curve) is compared to the three-term probability of 
\eqref{eq:Cervera-three-terms} (dashed curve).   Our formula predicts the probability to be somewhat lower 
throughout the $E_{\nu}$ region of peak oscillation.  This disparity is more pronounced at the longer baseline.  
}
\label{fig:1300-2450-km}
\end{figure}

To illustrate the level of improvement, we show the $\numu \rightarrow \nue$ probability 
for neutrinos of energies between 1.0 and 10 GeV,  propagating through the Earth 
for three baselines of interest to future experimentation.   The nominal value reported by the Daya Bay experiment, 
$\sin ^2 2\theta_{13} = 0.092$ \cite{ref:Daya-Bay},  is used throughout, and the normal mass hierarchy is everywhere assumed.  
Figure \ref{fig:1300-2450-km}a shows the $\nue$ appearance probability for the Fermilab to Homestake baseline of 1300 km 
as envisaged for the Long Baseline Neutrino Experiment (LBNE).    
At this baseline neutrino propagation is entirely through the terrestrial crust, 
and a uniform density of  2.72 g/cm$^3$ is assumed.  Our six-term formula (solid curve) 
agrees with the more approximate probability (dashed curve) fairly well, however 
a small reduction in $\nue$ appearance is indicated throughout the peak oscillation region.   
The difference between the predictions becomes negligible at shorter terrestrial baselines.

For baselines longer than LBNE
the reduction in $\nue$ appearance probability  
becomes more pronounced.    Figure \ref{fig:1300-2450-km}b compares expectations 
at the bimagic baseline of 2540 km.    At this baseline, there is propagation 
through the Earth's mantle as well as the crust, giving rise to a mean density 
of 3.2 g/cm$^3$.    In the vicinity of the oscillation peak at $\sim 4$ GeV our result 
falls below the Cervera {\it et al.} probability by several percent; this trend that persists at higher energies.
The same general trend has been shown by other improved approximation forms -- see for example,
Fig. 3 of Ref.~\cite{ref:Asano-Minakata}.

\begin{figure}[!htb]
\includegraphics[width=8.3cm]{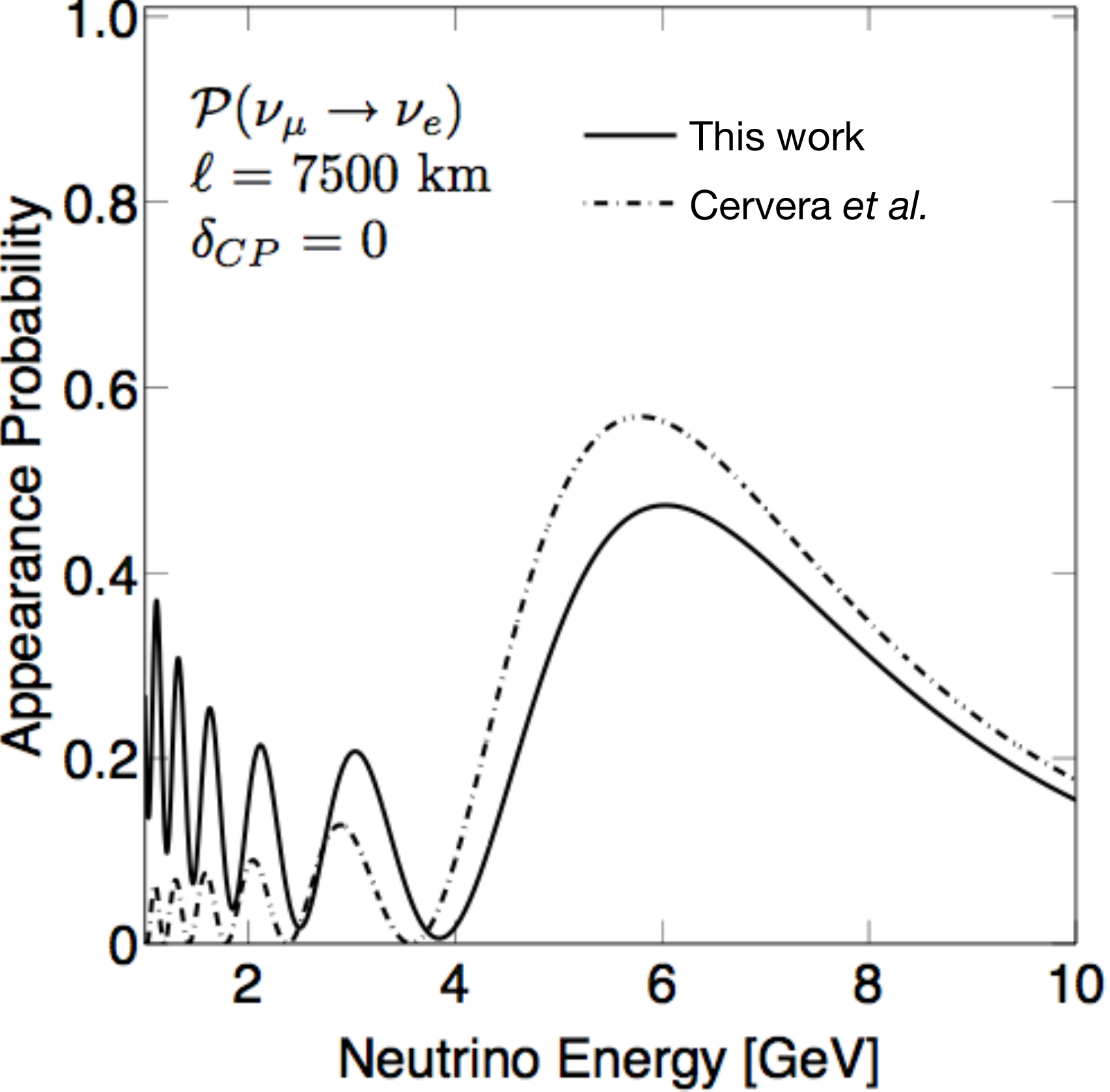}
\caption{ \small  Probability for $\numu \rightarrow \nue$ oscillation at the magic 7500 km baseline through the terrestrial mantle.    Expectations for our six-term formula (solid) and for the Cervera {\it et al.} result (dashed) are shown; a mean density of 4.3 gm/cm$^{3}$ is used for both.    The effect of the MSW resonance in the Earth's mantle is apparent in both curves;  our formula gives a probability which is lower by $\sim 8\%$ in the vicinity of the first oscillation peak.}
\label{fig:7500-km}
\end{figure}

With even longer baselines, the MSW resonance in the mantle~\cite{ref:MSW} greatly enhances 
the $\numu \rightarrow \nue$ oscillation.   Figure  \ref{fig:7500-km} compares 
Eqs.  (\ref{eq:probability}) and  \eqref{eq:Cervera-three-terms} at the so-called
magic baseline which occurs for propagation paths in the neighborhood of 7500 km.   
Here the propagation is predominantly mantle traversal and the mean density is 4.3 g/cm$^3$.   
The resonance-driven appearance probability is nearly $50\%$ at its peak, 
consequently Fig. \ref{fig:7500-km} is plotted with a distinctly larger abscissa range 
than is used in Figs. \ref{fig:1300-2450-km}a,b.  
Our probability formula exhibits the same shape as predicted by Eq. \eqref{eq:Cervera-three-terms} over most of
the $E_{\nu}$ range.   However it shows the appearance probability to be overestimated by Cervera {\it et al.}
throughout the region of the main oscillation peak.  
Figures \ref{fig:1300-2450-km}  and  \ref{fig:7500-km} indicate the extent to which 
the six-term probability of this work may offer improved accuracy for long baseline oscillations.

\begin{figure}[!htb]
\includegraphics[width=8.2cm]{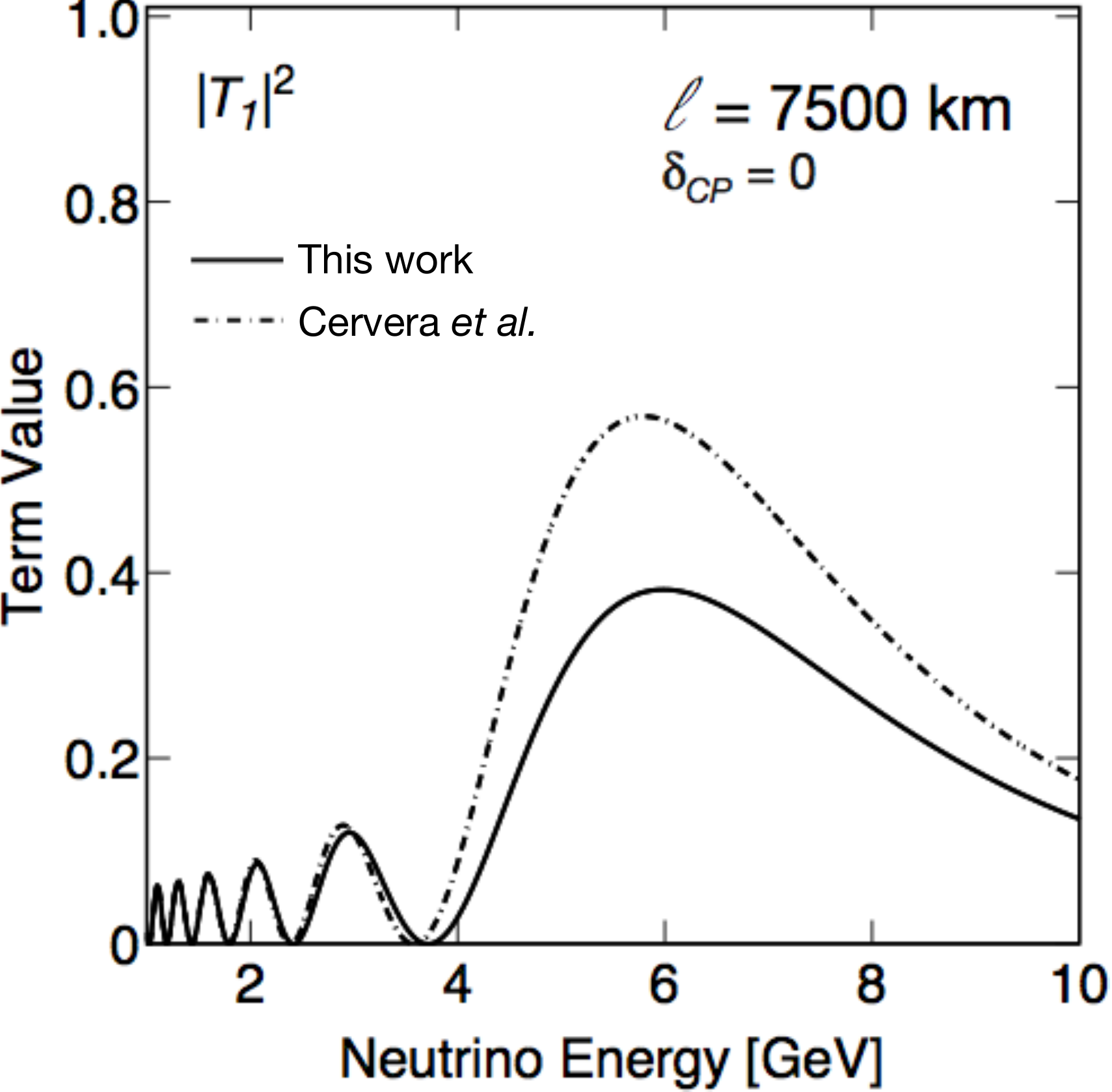}
\caption{
\small Comparison of the leading term in $\mathcal{P}(\numu \rightarrow \nue)$ from the formula of this
work (Eq. \eqref{eq:T_1_squared}) to the corresponding term in Eq. \eqref{eq:Cervera-three-terms}, 
for the 7500 km baseline.   the leading terms exhibit similar shape dependence with respect to neutrino energy, however $|T_{1}|^{2}$ of our formulation
gives a lower probability through the region of the MSW resonance.
}
\label{fig:7500-km-T1-sq}
\end{figure}

Figure \ref{fig:7500-km-T1-sq} compares $|T_1|^{2}$ of  Eq.  \eqref{eq:T_1_squared}, 
with the lead term of Eq. \eqref{eq:Cervera-three-terms}.    For the Cervera {\it et al.} result , 
the leading term gives the entire probability contribution at 7500 km whereas the second and third terms, 
which carry the CP-violating phase and parameters from solar-scale mixing, 
have negligible probability at the magic baseline.

Figure \ref{fig:7500-km-T1-sq} shows our $|T_1|^{2}$ term corresponds to a probability which is nearly $20\%$ 
below the Cervera {\it et al.} prediction in the maximum oscillation region.
This disagreement is partially alleviated by modest contributions, mostly of positive sign,  
which in our formula arise from the terms $( T_1 T_2^{*} + T_1^* T_2 )$ and $|T_2|^{2}$.     
The three other terms in our formula, namely those which involve 
 the $T_3$ amplitude, contribute an amount
which is only $\sim -0.002$  throughout the interval 2.0 $\le E_{\nu} \le 10.0$ GeV.

\section{Derivation of $\mathcal{A}(\nu_\mu \rightarrow \nu_e)$}
\label{sec:Derivation}
\subsection{Hamiltonian in flavor basis}

For neutrino propagation in vacuum, the Hamiltonian in the basis of three mass eigenstates
$\nu_i$ ($i$ =1,~2,~3) is $\hat H_0^{(i)} = \mathrm{diag}\left( 0, \frac{\Delta m_{21}^2}{2E}, \frac{\Delta m_{31}^2}{2E} \right\}
 = \frac{1}{2\ell_{v}} \cdot \mathrm{diag}   \left( 0, \alpha , 1\right) .$
The transformation from mass basis $\{\ket{\nu_i}\}$ to neutrino flavor basis $\{\ket{\nu_\varphi}\}$ 
($\varphi=e,~\mu,~\tau$) is provided by the unitary mixing matrix
\begin{equation} \label{eq:mixing_matrix_def}
  \vec \nu^{(\varphi)} = \hat U_{(mix)}~\vec \nu^{(i)} .
\end{equation}

\noindent
The Dirac-type CP-violating phase, $\delta_{CP}$, can be conveniently incorporated 
into $\hat U_{(mix)}$ by including auxiliary matrices $\mathbb{\hat I}_{\delta_{CP}}$ 
within the standard factored form:
\begin{equation} \label{eq:mixing_matrix}
  \hat U_{(mix)} \equiv \hat R_1(\theta_{23}) \cdot \mathbb{\hat I}_{\delta_{CP}} \cdot \hat R_2(\theta_{13}) \cdot 
  \mathbb{\hat I}_{-\delta_{CP}} \cdot \hat R_3(\theta_{12})
\end{equation}
where
$\mathbb{\hat I}_{\delta_{CP}} \equiv \mathrm{diag}(1, 1, e^{i\delta_{CP}})$ and 
$\mathbb{\hat I}_{\delta_{CP}}^\dagger = \mathbb{\hat I}_{-\delta_{CP}} = \mathrm{diag}(1, 1, e^{-i\delta_{CP}})$.
In Eq. (\ref{eq:mixing_matrix}) the atmospheric and solar mixings are accounted for via the rotation matrices $\hat R_1(\theta_{23})$ and $\hat R_3(\theta_{12})$, and the product $\mathbb{\hat I}_{\delta_{CP}} \cdot \hat R_2(\theta_{13}) \cdot \mathbb{\hat I}_{-\delta_{CP}}$ carries the CP-violating phase:
\[ \begin{split}
  \mathbb{\hat I}_{\delta_{CP}} \cdot \hat R_2 \cdot \mathbb{\hat I}_{-\delta_{CP}} =
    \left( \begin{array}{ccc}
                    c_{13} & 0 & s_{13} e^{-i\delta_{CP}} \\
                         0 & 1 & 0 \\
       -s_{13} e^{i\delta_{CP}} & 0 & c_{13} \\
    \end{array} \right).
\end{split} \]

The effective wave equation for vacuum propagation of flavor states is
\begin{equation} \label{eq:schroedinger_vacuum}
  i \frac{d}{dt} \vec \nu^{(\varphi)}(t) = \hat H_0^{(\varphi)} \vec \nu^{(\varphi)}(t).
\end{equation}
Here, the vacuum Hamiltonian in flavor basis is given by the unitary transform of $\hat H_0^{(i)}$,
\begin{equation} \label{eq:flavor_hamiltonian}
  \hat H_0^{(\varphi)} = 
    \left( \hat R_1 \mathbb{\hat I}_{\delta_{CP}} \hat R_2 \mathbb{\hat I}_{-\delta_{CP}} \hat R_3 \right)
    \hat H_0^{(i)}
    \left( \hat R_3^T \mathbb{\hat I}_{\delta_{CP}} \hat R_2^T \mathbb{\hat I}_{-\delta_{CP}} \hat R_1^T \right)
\end{equation}
Since $\mathbb{\hat I}_{-\delta_{CP}} (\mathbb{\hat I}_{\delta_{CP}})$ commutes with $\hat R_3 (\hat R_3^T)$,
and since
$\mathbb{\hat I}_{-\delta_{CP}} \hat H_0^{(i)} \mathbb{\hat I}_{\delta_{CP}} = \hat H_0^{(i)}$,
Eq. (\ref{eq:flavor_hamiltonian}) can be simplified:
\begin{equation}\label{eq:hamiltonian_simplified_1}
\begin{split}
  \hat H_0^{(\varphi)}
    &= \left( \hat R_1 \mathbb{\hat I}_{\delta_{CP}} \right)
       \left( \hat R_2 \hat R_3 \hat H_0^{(i)} \hat R_3^T \hat R_2^T \right)
       \left( \mathbb{\hat I}_{-\delta_{CP}} \hat R_1^T \right) \\
    &= \left( \hat R_1 \mathbb{\hat I}_{\delta_{CP}} \right)
       \hat H_0^{(23)}
       \left( \mathbb{\hat I}_{-\delta_{CP}} \hat R_1^T \right) ~,
\end{split}
\end{equation} 
where
\begin{equation} \label{eq:hamiltonian_23-b} \nonumber
 \hat H_0^{(23)}  
    = \frac{1}{2\ell_v}
      \left( \begin{array}{ccc}           
           s_{12}^2c_{13}^2\alpha + s_{13}^2 &  \frac{1}{2} c_{13}\alpha'   &  \frac{1}{2} \sin2\tilde\theta_{13} \\
                        \frac{1}{2}  c_{13}\alpha' & c_{12}^2 \alpha &  -\frac{1}{2}  s_{13} \alpha' \\
        \frac{1}{2} \sin2\tilde\theta_{13}  &  -\frac{1}{2} s_{13}\alpha' &  s_{12}^2 s_{13}^2\alpha + c_{13}^2 \\            
      \end{array} \right) .   
\end{equation}
\noindent 
Upon inclusion of the MSW matter interaction 
\[ \begin{split}
  \hat H_{\mathrm{matter}}^{(\varphi)} = \mathrm{diag} \left( V_e, 0, 0 \right) ~,
\end{split} \]
\noindent
the total Hamiltonian including matter effects
as well as CP violation can be written in flavor basis as 
\begin{equation} \label{eq:flavor_hamiltonian_matter}
  \hat H^{(\varphi)} = 
    \left( \hat R_1 \mathbb{\hat I}_{\delta_{CP}} \right)
    \left( \hat H_0^{(23)} + \hat H_{\mathrm{matter}}^{(\varphi)} \right)
    \left( \mathbb{\hat I}_{-\delta_{CP}} \hat R_1^T \right).
\end{equation}
Thus the full wave equation in flavor basis is 
Eq. (\ref{eq:schroedinger_vacuum}) with $\hat H^{(\varphi)}$ defined by Eq. (\ref{eq:flavor_hamiltonian_matter})
replacing $\hat H_0^{(\varphi)}$ on the right-hand side.

\subsection{Hamiltonian in propagation basis}

Neutrino propagation is usefully re-cast by transforming to the propagation basis.  
The latter basis is defined via

\begin{equation} \label{eq:propagation_basis}
  \vec \nu^{(p)} = \mathbb{\hat  I}_{-\delta_{CP}} \hat R_1^T(\theta_{23}) \vec \nu^{(\varphi)}
  \;\ ,\;\,
  \vec \nu^{(\varphi)} = \hat R_1 \mathbb{\hat I}_{\delta_{CP}} \vec \nu^{(p)}.
\end{equation}
Multiplication of the wave equation in flavor basis from the left by 
$\mathbb{\hat I}_{-\delta_{CP}} \hat R_1^T$ yields
\begin{equation} \label{eq:propagation_schroedinger_form}
  i \frac{d}{dt} \vec \nu^{(p)}(t) = \hat H^{(p)}  \vec \nu^{(p)} ~,
\end{equation}
where $  \hat H^{(p)} \equiv \left( \hat H_0^{(23)} + \hat H_{\mathrm{matter}}^{(\varphi)} \right)$
is the effective Hamiltonian in the propagation basis.  
The matrix $\hat H^{(p)}$ is nearly identical to $ \hat H_0^{(23)}$ of Eq.~(\ref{eq:hamiltonian_23-b}) but includes
the matter term $V_e = A /2\ell_{\nu}$ added to the element  $(\hat H_0^{(23)})_{11}$.
Hamiltonian $\hat H^{(p)}$ 
is a real-valued, symmetric matrix devoid of the CP-violating phase $\delta_{CP}$.   Of course the CP-phase reappears when
one transforms from propagation basis back into flavor basis.

\vskip 5pt

The matrix $\hat H^{(p)}$ can be ``re-phased".  That is,
we perform an algebraic manipulation leading to 
removal of a term proportional to $\mathbb{\hat I}$, which merely contributes an overall phase to the oscillation amplitudes. 
Specifically we subtract (and also add, but then discard) 
the following diagonal matrices to $\hat H^{(p)}$:

\begin{equation} \label{eq:rephasings}
\begin{split}
 & \frac{c_{13}^2}{2\ell_v} \mathbb{\hat I} ~,  ~~~~\frac{1}{2} \cdot \frac{1}{2\ell_v} (A - \cos 2\theta_{13} ) \mathbb{\hat I} ~, ~~\text{and}~~ \\
 & \frac{s_{12}^2 \alpha}{2 \ell_v} \left( \frac{c_{13}^2}{2} + \frac{s_{13}^2}{2} \right) = \frac{1}{4 \ell_v} s_{12}^2 \alpha.
\end{split} 
\end{equation}
Upon extraction of a factor $\frac{1}{2}$ we obtain
\begin{equation} \label{eq:full_hamiltonian_prop_3} \nonumber
\begin{split}
  &\hat H^{(p)} = \frac{1}{4\ell_v} \cdot \\
& \left( \begin{array}{ccc}     
  - ( \cos2\tilde{\theta}_{13} - A)  & c_{13} \alpha'       &  \sin2\tilde{\theta}_{13}     \\
  c_{13} \alpha'       & -\left[ (1 + A) + \alpha'' \right]     &  -s_{13} \alpha'  \\
  \sin2\tilde{\theta}_{13}   &  -s_{13} \alpha'    &    + ( \cos2\tilde{\theta}_{13} - A)  \\                    
      \end{array} \right) .   
\end{split}
\end{equation}

\noindent
Using the variable G defined in Eq. \eqref{eq:Define-eta_and_G}, we identify $\hat H^{(p)}_{22} = -G$.
To represent the other elements of $\hat H^{(p)}$ in a compact form, we define:
\begin{equation} \label{eq:symbols}
\begin{split}
  &Q \equiv \frac{1}{4\ell_v}(\cos2\tilde{\theta}_{13} - A) ,  ~~ 
    f \equiv  \frac{1}{4\ell_v}  \sin2\tilde{\theta}_{13}  , \\
  &a \equiv \frac{1}{4\ell_v}  \left[   c_{13} \alpha'   \right]  ,  ~~
   b \equiv \frac{1}{4\ell_v}  \left[   - s_{13}  \alpha'   \right] . 
\end{split} 
\end{equation} 
The full Hamiltonian in propagation basis can then be written as
\begin{equation} \label{eq:hamiltonian_propagation_simplest}
  \hat H^{(p)}
    = \left( \begin{array}{ccc}
      - Q & a & f \\
      a & -G & b \\
      f & b & + Q 
    \end{array} \right).
\end{equation}

\subsection{Formulation in an interaction picture}

We separate 
$\hat H^{(p)}$ into an ``unperturbed" part, $\hat H_0^{(p)}$, plus an interaction potential, $\mathbb{\hat V}$ comprised of elements proportional to $\alpha'$:
\begin{equation} \label{eq:full_hamiltonian_prop_base}
  \hat H^{(p)} =  \hat H_0^{(p)} +  \mathbb{\hat V} = 
  \left(\begin{array}{ccc}
    -Q &  0 &  f \\
     0 & -G &  0 \\
     f &  0 & +Q
  \end{array}\right) + 
  \left(\begin{array}{ccc}
     0 &  a &  0 \\
     a &  0 &  b \\
     0 &  b & 0
  \end{array}\right).
\end{equation}
Wave equation \eqref{eq:propagation_schroedinger_form} can then be re-cast into an Interaction
picture:
\begin{equation} \label{eq:interaction}
  \vec \nu^{(I)}(t) = e^{i \hat H_0^{(p)}t} \vec \nu^{(p)}(t),\,\,
  \vec \nu^{(p)}(t) = e^{-i \hat H_0^{(p)}t} \vec \nu^{(I)}(t)  ~,
\end{equation}
so that
\begin{equation} \label{eq:state_evolution_inter_2}
  i \frac{d}{dt} \vec \nu^{(I)}(t) 
    =  \hat V_I \cdot \vec \nu^{(I)} (t)
\end{equation}
where
\begin{equation} \label{eq:V_I}
  \hat V_I(t) = e^{i \hat H_0^{(p)}t} \cdot \mathbb{\hat V} \cdot e^{-i \hat H_0^{(p)}t}.
\end{equation}

\vskip 5pt

Our approach to Eq. \eqref{eq:state_evolution_inter_2} is to construct a heuristic
functional form $\hat U_I^{(est)}(t,\,0)$ which serves as an estimator of the time evolution operator in the
Interaction picture:
\begin{equation} \label{eq:interaction_time}
  \vec \nu^{(I)}(t) \simeq \hat U_I^{(est)}(t,\,0)  \cdot \vec \nu^{(I)}(0).
\end{equation}
The wave equation which defines our $ \hat U_I^{(est)}(t,\,0)$ is
\begin{equation} \label{eq:wave_eq_time_evolve_op}
  i \frac{d}{dt} \hat U_I ^{(est)}(t, 0) = \hat V_I (\ell) \cdot \hat U_I ^{(est)}(t, 0).
\end{equation}
where $t$ is variable but $\hat V_I(t)$ is set to its value at the final baseline distance $t = \ell$.

To obtain $\hat V_I(t)$ we require the matrix representation (in propagation basis) of the unitary
operator forms $\mathrm{exp}(\pm i \hat H_0^{(p)} t)$, where $\hat H_0^{(p)}$
has the elements given in Eq. \eqref{eq:full_hamiltonian_prop_base}.
\noindent
Considering the  series expansion
\begin{equation} \label{eq:exp_sum}
  \hat W \equiv e^{i \hat H_0^{(p)} t}
    = \sum_{n=0}^{\infty} \frac{\left(i \hat H_0^{(p)} t \right)^n}{n!},
\end{equation}
it is readily seen that neither the middle row nor middle column of matrix $ \hat H_0^{(p)}$ 
mixes with other elements.
Then the matrix $\hat W$ has the form
\begin{equation} \label{eq:exp_matrix}
  e^{i \hat H_0^{(p)} t}
    = \left( \begin{array}{ccc}
        W_{11} &           0 & W_{13} \\
             0 & e^{-iG t} &      0 \\
        W_{31} &      0      & W_{33}
      \end{array} \right)~.
\end{equation}
Thus we may work with the reduced $2\times2$ matrix
\begin{equation} \label{eq:reduced_matrix}
  \hat H_{R}^{(p)} = \left( \begin{array}{cc} -Q & f \\ f & +Q \end{array} \right) 
      = f \hat \sigma_x - Q \hat \sigma_z
\end{equation}
where $\hat \sigma_{x, z}$ are the Pauli spinor matrices. We write $\hat H_{R}^{(p)} = \vec N \cdot \vec \sigma$ ,
where $\vec N = (f, 0, -Q)$, with $  | \vec N |  = \sqrt{f^2 + Q^2} $  (see Eq. \eqref{eq:Define-N}).
The unit vector $\hat n \equiv \vec N / N = \left( f^2 + Q^2 \right)^{-1/2}(f, 0, -Q)$
serves as the axis-of-rotation in the reduced spinor space. 
With $\hat H_{R}^{(p)} = N \hat n \cdot \vec \sigma$, and recognizing that $ t = \ell$ in natural units,  we write
\begin{equation} \label{eq:rotation}
  e^{i \hat H_{R}^{(p)}(t=\ell)}
    = e^{i \hat n \cdot \vec \sigma(N\ell)}
    = e^{i \hat n \cdot \vec \sigma \phi} ~,
\end{equation}
where $\phi \equiv N \ell$ designates the rotation angle.
With $\hat n = (n_{x}, 0 , n_{z})$, the spinor identity is
\begin{equation} \label{eq:spinor_identity}
\begin{split}
  e^{i \hat n \cdot \vec \sigma \phi}
    &= \mathbb{\hat I} \cos \phi + i \vec \sigma \cdot \hat n \sin \phi \\
    &= \left(\begin{array}{cc}
         \cos\phi + i n_z \sin\phi & i n_x \sin\phi \\
         i n_x \sin\phi            & \cos\phi - i n_z \sin\phi
       \end{array}\right) .
\end{split}
\end{equation}
 We define $\gamma \equiv \cos\phi + i n_z \sin\phi$ and
$\beta \equiv n_x \sin\phi$,  and we write
\begin{equation} \label{eq:spinor_identity_final}
\begin{split}
  e^{i \hat H_0^{(p)} \ell}
    &= \left(\begin{array}{ccc}
         \gamma & 0 & i \beta \\
         0 & e^{-i G\ell} & 0 \\
         i\beta & 0  & \gamma^*
       \end{array}\right).
\end{split}
\end{equation}
Note that $n_x$ and $n_z$ are real-valued, hence $\beta$ is
real-valued, however $\gamma$ is complex.  
Evaluation of Eq. \eqref{eq:V_I} yields
\begin{equation} \label{eq:V_I_2}
\begin{split}
  &\hat V_I(\ell) = \\
    &\left(\begin{array}{ccc}
                             0 & (\gamma a + i \beta b) e^{iG\ell} &                    0 \\
    (\gamma^{*} a -i \beta b) e^{-iG\ell} &                     0 & ( \gamma b - i \beta a) e^{-iG\ell} \\
                             0 & (\gamma^* b + i \beta a ) e^{iG\ell} &                    0
       \end{array}\right).
\end{split}
\end{equation}
The complex matrix elements of \eqref{eq:V_I_2} are usefully expressed as
\begin{equation} \label{eq:u_v_def}
\begin{split}
  u &\equiv (\gamma a + i \beta b) e^{iG\ell} , \\
  v &\equiv ( \gamma b - i \beta a) e^{-iG\ell} .
\end{split}
\end{equation}
Then we have
\begin{equation} \label{eq:V_I_final}
  \hat V_I (\ell)
     = \left(\begin{array}{ccc}
         0  & u & 0 \\
         u^* & 0 & v \\
         0 & v^* & 0
       \end{array}\right).
\end{equation}
where $|u|^2 + |v|^2 = a^2 + b^2 = (\alpha ' /4\ell_v)^2 = \eta^2$.

\subsection{Heuristic construction for $ \hat U_I (t,\,0)$}

Exact solution of wave equation \eqref{eq:state_evolution_inter_2} requires a time evolution
operator $ \hat U_I (t,\,0)$ which solves Eq. \eqref{eq:wave_eq_time_evolve_op} for the case
wherein $ \hat V_I (t)$ is a function of ``live" variable $t$.    A formal solution is provided in
principle by the Dyson series.    In practice, the series is always truncated at low order;  Sec. III of
Ref. \cite{ref:Asano-Minakata} gives a clear discussion.    As an alternative approximation which
retains the perturbative expansion structure of the Dyson series, we introduce the exponentiation
of $ \hat V_I (\ell)$ as a heuristic form:
\begin{equation} \label{eq:U_I}
  \hat U_I ^{(est)}(t = \ell,0) = e^{-i \hat V_I (\ell)t} .
\end{equation}
To obtain the matrix representation of $ \hat U_I ^{(est)}(\ell,0)$ we take a 
brute force approach, rather than e.g.,  harnessing the Cayley-Hamilton theorem \cite{ref:Ohlsson-JMP}.   We observe that
\begin{equation} \label{eq:Theta_def}
  \left( \hat V_I \right)^2
    = \left(\begin{array}{ccc}
         |u|^2 &           0 & uv   \\
             0 & |u|^2+|v|^2 &  0    \\
        (uv)^* &           0 & |v|^2
      \end{array}\right) ~~.
\end{equation}
Furthermore, $\left( \hat V_I \right)^3 = \eta^2 \left(\hat V_I\right)$,
$\left( \hat V_I \right)^4 = \eta^2 \left( \hat V_I \right)^2$ , and by induction  $\left( \hat V_I \right)^{n=\mathrm{odd}}= \eta^{n-1} \hat V_I \mathrm{,\;and} ~
  \left( \hat V_I \right)^{n=\mathrm{even}}= \eta^{n-2} \left( \hat V_I \right)^2$.
Thus
\begin{equation} \label{eq:exp_expanded}
\begin{split}
  e^{-i \hat V_I \ell}
    &= \sum_{n=0}^{\infty} \frac{\left( -i \hat V_I \ell \right)^n}{n!} \\
    &= \mathbb{\hat I} - \left(\frac{\hat V_I}{\eta}\right)^2\left(1 - \cos \eta \ell \right)
      - i \frac{\hat V_I}{\eta} \sin \eta \ell.
\end{split}
\end{equation}

\noindent
The time evolution operator for neutrino propagation
in our Interaction picture obeys a matrix identity reminiscent of that
for rotations generated by $\hat J_y^{(j=1)}$ \cite{ref:Sakurai}.

\vskip 2pt

Equation \eqref{eq:exp_expanded} yields an explicit representation for $e^{-i \hat V_I \ell}$.
We write 
$ \theta \equiv \eta \ell  , ~~  \overline{u} \equiv u/\eta, ~~ \overline{v} \equiv v/\eta $, and use
$(1 - \cos\theta ) = 2\cdot  \sin^2 \frac{\theta}{2} $.   Then the evolution operator of Eq. \eqref{eq:U_I} is
\begin{equation} \label{eq:V_explicit}
\begin{split}
   &\hat U_I^{(est)}(\ell, 0) = \\
    & \left( \begin{array}{ccc}
         1 - 2 |\overline{u}|^2 \sin^2\frac{\theta}{2} &
         -i \overline{u} \sin \theta &
          - 2 \overline{u} \overline{v} \sin^2\frac{\theta}{2} \\
          -i \overline{u}^* \sin \theta &
            \cos \theta &
           -i \overline{v} \sin \theta  \\
         - 2 (\overline{u} \overline{v})^* \sin^2\frac{\theta}{2} &
          -i \overline{v}^* \sin \theta   &
         1 - 2 |\overline{v}|^2 \sin^2\frac{\theta}{2}  
       \end{array} \right) .
       \end{split}
\end{equation}

We proceed to work our way back, first to the three-neutrino propagation
basis and then to the neutrino flavor basis, wherein the matrix
elements of $\hat U^{(\varphi)}(\ell,0)$ correspond to the possible neutrino
oscillation amplitudes for 3-flavor mixing.
The return of the time evolution operator to propagation basis requires
that we calculate
\begin{equation} \label{eq:v_prop}
  \hat U^{(p)}(\ell,0) = e^{-i \hat H_0^{(p)} \ell} \cdot \hat U_I^{(est)}(\ell,0) .
\end{equation}
The required matrix forms are \eqref{eq:V_explicit} and the adjoint of
\eqref{eq:spinor_identity_final}. As a preliminary to this multiplication,
we define some compact forms.   For the diagonal elements we define
\begin{equation} \label{eq:func_forms-diagonal}
\begin{split}
  &D_{u} \equiv 1 - 2 |\overline{u}|^2 \sin^2\frac{\theta}{2}, ~~
  d \equiv \cos \theta , ~~ \\
  &D_{v} \equiv 1 - 2 |\overline{v}|^2 \sin^2\frac{\theta}{2}.  
\end{split}
\end{equation}
For the off diagonal elements, we define
\begin{equation} \label{eq:func_forms-off-diag}
 e \equiv \overline{u} \sin \theta, ~~ p \equiv -2 \overline{u}\overline{v} \sin^2\frac{\theta}{2}, ~~ k \equiv \overline{v} \sin \theta .
\end{equation}
Then we write
\begin{equation} \label{eq:V_explicit_simple}
\begin{split}
  \hat U_I^{(est)}(\ell, 0)
    &= \left( \begin{array}{ccc}
         D_{u}  & -i e   & p   \\
          -i e^*    & d & -i k   \\
         p^* & -i k^*& D_{v}
       \end{array} \right)~.
\end{split}
\end{equation}
Proceeding with the evaluation of \eqref{eq:v_prop}:
\begin{equation} \label{eq:V_p_explicit} \nonumber
\begin{split}
  &\hat U^{(p)}(\ell, 0)
    = \\
&\left( \begin{array}{ccc}
         (\gamma^* D_{u} - i \beta p^*) &
        (\gamma^* (-ie) -  \beta k^*) &
         (\gamma^* p - i \beta D_{v}) \\
         (-ie^* )~e^{iG\ell} &
          d~e^{iG\ell} &
         (-ik)~e^{iG\ell} \\
          (\gamma p^*  -i \beta D_{u}) &
          (\gamma (-i k^*)  - \beta e) &
          ( \gamma D_{v}  -i \beta p)        
       \end{array} \right).
\end{split}
\end{equation}
Finally, we return to neutrino flavor basis via
\begin{equation} \label{eq:V_neutrino_basis} \nonumber
  \hat U^{(\varphi)}(\ell,0) =
  \hat R_1(\theta_{23}) \cdot \mathbb{\hat I}_{\delta_{CP}} \cdot \hat U^{(p)}(\ell,0) \cdot
  \mathbb{\hat I}_{-\delta_{CP}} \cdot \hat R_1^T(\theta_{23})  .
\end{equation}
\smallskip
All of the physically relevant neutrino flavor oscillation amplitudes are contained
in the matrix $\hat U^{(\varphi)}(\ell,0)$;  however our focus here is upon the element $U^{(\varphi)}_{12}$.

\subsection{$\nu_\mu \rightarrow \nu_e$ oscillation amplitude}

Element $U^{(\varphi)}_{12}$ provides the $\nu_e$ appearance amplitude from an initial beam of
$\nu_\mu$ neutrinos:

\begin{equation}
  U^{(\varphi)}_{12}
    = \mathcal{A}(\nu_\mu \rightarrow \nu_e)
    = c_{23} U^{(p)}_{12} + s_{23} U^{(p)}_{13} e^{-i \delta_{CP}}.
\end{equation}

Unwinding notations and rearranging, we obtain

\begin{equation}
\begin{split} \label{eq:amplitude_exp}
  \mathcal{A}(\nu_\mu \rightarrow \nu_e)
    = & ~(-i) s_{23} \beta e^{-i\delta_{CP}} \\
      & + (-i) c_{23} \left[ \gamma^* \bar u - i \beta \bar v^* \right] \cdot \sin\theta \\
      & + 2 s_{23} \left[ i \beta |\bar v|^2 - \gamma^* \bar u \bar v \right]
        \cdot \sin^2 \frac{\theta}{2} \cdot e^{-i\delta_{CP}}.
\end{split}
\end{equation}

We identify the three terms of Eq. (\ref{eq:amplitude_exp}) using the 
expression $\mathcal{A}(\nu_\mu \rightarrow \nu_e) = T_1 + T_2 + T_3$.  
For $T_1$ we have
\begin{equation}
\begin{split}
  T_1 &= (-i) s_{23} \beta e^{-i\delta_{CP}} = (-i) s_{23} n_x \sin \phi \cdot e^{-i\delta_{CP}} \\
      &= (-i)\sin 2\tilde\theta_{13}~ s_{23} \cdot \frac{\sin\phi}{4 \ell_v N} \cdot e^{-i\delta_{CP}},
\end{split}
\end{equation}
which coincides with Eq. \eqref{eq:T_1_final}.

For $T_2$ we have
\begin{equation}
\begin{split}
  T_2 = -i c_{23} \left[ \gamma^* \bar u - i\beta \bar v^* \right] \sin\theta ~~.
\end{split}
\end{equation}

Assembling the various factors,
\begin{equation}\nonumber
\begin{split}
  T_2 &= -i \frac{c_{23}}{\eta}
         \left[ \gamma^*(\gamma a + i \beta b) - i\beta \gamma^* b + \beta^2 a \right]\cdot
         \sin\theta\cdot  e^{iG\ell}, \\
      &= -i c_{23} \cdot \frac{a}{\eta} \cdot \sin(\theta) \cdot e^{iG\ell},
\end{split}
\end{equation}
which is the same as Eq. \eqref{eq:T_2_final}.

For the remaining term $T_3$ we have
\begin{equation} \label{eq:T_3}
\begin{split}
  T_3 = 2 s_{23} \left[ i \beta |\bar v|^2 - \gamma^* \bar u \bar v \right]
        \cdot \sin^2 \frac{\theta}{2} \cdot e^{-i\delta_{CP}}.
\end{split}
\end{equation}

Reduction within the bracket leads to
\begin{equation}
\begin{split} \label{eq:T_3_2}
  T_3 = -2 s_{23} \left[
     \gamma \left(\frac{ab}{\eta^2}\right) - i\beta \frac{a^2}{\eta^2} \right]
     \cdot \sin^2 \frac{\theta}{2} \cdot e^{-i\delta_{CP}}.
\end{split}
\end{equation}
Now $ab/\eta^{2} = - c_{13}s_{13}$ and $a^2/\eta^2 = c_{13}^2$.   Recall that $ \gamma = \cos\phi + i n_z \sin\phi,$ and that
$ \beta  = n_x \sin\phi = \frac{f}{N}\sin\phi = \sin 2\tilde\theta_{13} \cdot  \frac{ \sin\phi }{4\ell_v N}.$
Working within the bracket of Eq. (\ref{eq:T_3_2}), we separate
the real and imaginary pieces:

\begin{equation}\nonumber
\begin{split}
   T_3 = &2 s_{23} \left\{ \frac{\sin 2 \theta_{13}}{2}\cos\phi
             +i\left[ \frac{\sin 2 \theta_{13}}{2} n_z + c_{13}^2 n_x \right] \sin\phi \right\}  \cdot \\
              &\sin^2 \left( \frac{\eta \ell}{2} \right) \cdot e^{-i\delta_{CP}}.
\end{split}
 \end{equation}
Then
\begin{equation}\nonumber
\begin{split}
  T_3 = &\sin 2\theta_{13} ~s_{23} \cdot  \left[ \cos(N\ell) + iF_{A} \frac{ \sin(N\ell)}{4 \ell_v N} \right] \cdot \\
        &\sin^2 \left( \frac{\eta \ell}{2} \right) \cdot e^{-i\delta_{CP}}
\end{split}
\end{equation}
which is identical to Eq. \eqref{eq:T_3_final}.

\vspace{9pt}
\section{Conclusion}
\label{sec:Conclusion}

Having presented and subsequently derived each of the probability terms individually, we
conclude by stating the entire six-term formula for $\mathcal{P}(\numu \rightarrow \nue)$ :
\begin{equation}  \label{eq:Entire-Exact-Formula}
\begin{split}
  \mathcal{P}&(\nu_\mu \rightarrow \nu_e) =  \\
& ~~~~~~~(\sin 2 \tilde\theta_{13})^2 ~s_{23}^2 \cdot \frac{  \sin^2(D \Delta)}{D^2} \\
&~~~~~+ \sin 2 \tilde\theta_{13} ~c_{13}\sin 2\theta_{23}   
       \cdot \sin(\alpha'  \Delta)  \cdot \frac{\sin(D \Delta)}{D} ~\cdot \\
&~~~~~~~~~~~~[ \cos \Delta' \cdot \cos \delta_{CP} - \sin \Delta' \cdot \sin \delta_{CP} ]  \\
&~~~~~ +  c_{13}^2 c_{23}^2 \cdot \sin^2 (\alpha'  \Delta) \\
 &~-2  \sin 2 \theta_{13}  \sin 2 \tilde \theta_{13}  s_{23}^2 ~ F_A 
      \cdot \sin^2 \left(\frac{\alpha' \Delta}{2} \right) \cdot \frac{ \sin^2(D \Delta)}{D^2} \\ 
 &~+ \sin 2\theta_{13} ~c_{13} \sin 2 \theta_{23}  \cdot \sin(\alpha'  \Delta) \cdot
       \sin^2 \left( \frac{\alpha' \Delta}{2} \right) \cdot \\
          &~\left\{ \cos(D \Delta) \cdot \sin(\Delta' + \delta_{CP})
         - F_A\frac{ \sin(D \Delta)}{D} \cdot \cos(\Delta'+\delta_{CP}) \right\} \\
 +&  \sin^2 2\theta_{13} ~s_{23}^2\cdot \sin^4 \left(\frac{\alpha'  \Delta}{2} \right) \cdot 
     \left\{ \cos^2(D \Delta) + F_A^2  \cdot  \frac{\sin^2(D \Delta)}{D^2} \right\}. 
\end{split}
\end{equation}
As discussed in Sec. \ref{sec:Comparison}, the three leading terms of Eq. \eqref{eq:Entire-Exact-Formula} 
are reminiscent of the perturbative expansion of Eq. \eqref{eq:Cervera-three-terms}.
These three terms account for essentially all of the $\numu \rightarrow \nue$ appearance probability.

\vspace{+7pt}
\section{Acknowledgments} \vspace{-8pt}

This work was supported by the United States Department of Energy under grant DE-FG02-92ER40702.

\end{document}